\documentclass[prd,aps]{revtex4}

\usepackage{epsfig}

\newcommand{\beq}{\begin{equation}}
\newcommand{\eeq}{\end{equation}}
\newcommand{\be}{\begin{eqnarray}}
\newcommand{\ee}{\end{eqnarray}}

\begin{document}

%\begin{titlepage}

%%\voffset 1.5cm
%\draft
\preprint{
KAIST-TH 2004/08
%\begin{tabular}{l}
%\hbox to\hsize{\hfill KIAS-P03046}\\[-2mm]
%\hbox to\hsize{           \hfill SNUTP 03-013}\\[-3mm]
%\end{tabular}
}

\title{
Time-dependent CP asymmetry in $B^0 \to \rho^0 \gamma$ decay \\
to probe the origin of CP violation}

\author{
C. S. Kim$^{1,2}$\thanks{cskim@yonsei.ac.kr},~~
Yeong Gyun Kim$^3$\thanks{yg-kim@korea.ac.kr},~~
Kang Young Lee$^4$\thanks{kylee@muon.kaist.ac.kr}
}
\vskip 0.5cm

\affiliation{
$^1$ Department of Physics,
Yonsei University, Seoul 120-749, Korea\\
$^2$ Physics Division, National Center for Theoretical Sciences,
Hsinchu 300, Taiwan\\
$^3$ Department of Physics,
Korea University, Seoul 136-701, Korea\\
$^4$ Department of Physics,
Korea Advanced Institute of Science and Technology,
Daejeon 305-701, Korea
}
\vskip 0.5cm

\date{\today}
%\vspace{2cm}
%\maketitle

\begin{abstract}
Since the CP violation in the $B$ system has been investigated
up to now only through processes related to the $B$--$\bar{B}$ mixing,
urgently required is new way of study for the CP violation
and establishing its origin in the $B$ system
%measuring $\beta$,
independent of the mixing process.
In this work, we explore the exclusive $ B^0 \to \rho^0 \gamma$ decay
to obtain the time-dependent CP asymmetry in $b \to d$ decay process
in the standard model and the supersymmetric model.
We find that the complex RL and RR mass insertion
to the squark sector in the MSSM
can lead to a large CP asymmetry in $b \to d \gamma$ decay
through the gluino-squark diagrams,
which is not predicted in the Standard Model
induced by the $B$--$\bar{B}$ mixing.

\end{abstract}


\maketitle

%\end{titlepage}

%\voffset 0cm

%\begin{multicols}{2}
%\narrowtext

%\tightenlines

%\newpage

\section{Introduction}

The unitarity of the Cabibbo-Kobayashi-Maskawa (CKM) matrix
yields the relation $V_{ud}V_{ub}^*+V_{cd}V_{cb}^*+V_{td}V_{tb}^*=0$,
which describes a triangle on the complex plane.
The angles of this triangle are defined by
\be
\alpha~ (\phi_2) \equiv
{\rm arg}\left(- \frac{V_{td} V_{tb}^*}{V_{ud} V_{ub}^*} \right),~~~~
\beta~ (\phi_1) \equiv
{\rm arg}\left(- \frac{V_{cd} V_{cb}^*}{V_{td} V_{tb}^*} \right),~~~~
\gamma~ (\phi_3) \equiv
{\rm arg}\left(- \frac{V_{ud} V_{ub}^*}{V_{cd} V_{cb}^*} \right),
\ee
of which non-zero values indicate the existence of the  CP violation.
In the $B$ meson system, the CP asymmetry in  $B \to J/\psi K_S$ decays,
which is directly related to $\sin 2 \beta$
through $B^0$--$\bar{B}^0$ mixing,
has been measured and agrees well
with the standard model (SM) prediction by the CKM unitarity.
Since the present CP violating asymmetry appears
via $\beta$, presently measured only through processes involving
the $B^0$--$\bar{B}^0$ mixing,
it is strongly required to find  new observables
of the CP violation in the $B$ system
and establish that the measured $\beta$ indeed originates
from $V_{td}$ of the CKM in a way independent of the mixing.
Within the SM the source of the CP violation of the $B$ system
(and the weak phase $\beta$) is nothing but
the phase of CKM matrix element $V_{td}$ in the leading order.
The magnitude and the phase of $V_{td}$ cannot be measured
in a direct manner due to top quark's fast decay
before its hadronization. This leaves only  indirect methods for
experimental determination in various processes.

As is well known, $\sin 2 \beta$
can be determined independently of the $B^0$--$\bar{B}^0$ mixing
in a theoretically clean manner,
up to $\Delta \sin 2 \beta \approx \pm 0.10$,
provided $B(K^+ \to \pi^+ \nu \bar\nu)$ and
$B(K_L \to \pi^0 \nu \bar\nu)$ are measured within $\pm 10 \%$ accuracy
\cite{GBAB}.
Why do we urgently require yet another method to examine
the CP violation in the $B$ system?
The main reason is that this area is very likely
to yield information about new physics beyond the SM.
We expect that new physics can influence the $\Delta B = 1$ penguin decays
in a different way than the usual $\Delta B =2$ mixing.
Examples are the {\it possibly} large new physics effects
recently measured in $b \to s \bar s s$ decays.
The measurements of $\sin 2 \beta$ in $b \to s \bar s$ decays
may have shown deviations from that in  $B \to J/\psi K_S$ decay,
which are still controversial \cite{sin2beta}.
A sizable difference between them could be a clear
evidence of the new physics beyond the SM.
If one of such discrepancy indeed exists, it implies an evidence of a
new physics effect beyond the SM \cite{DKS}, and also suggests
an interesting possibility that the new physics effect is large
in the penguin diagrams while its contribution to
the $B^0$--$\bar{B}^0$ mixing is small.

The Cabibbo-suppressed $b \to d \gamma$ decay also involves $V_{td}$
and provides us a new chance to measure the weak phase $\beta$
in a way independent from the mixing
to study the CP violation in the $B$ system.
Although the inclusive  $B \to X_d \gamma$ decay is a
theoretically clean process to determine $V_{td}$ \cite{KSM},
it can be hardly discriminated from large $B \to X_s \gamma$ background
in the experiment.
Thus our interest is devoted to the study of the exclusive channels,
$B \to \rho \gamma$ and $B \to \omega \gamma$.
The first measurement on the branching ratio
of $ b \to d \gamma$ process,
$$
{\rm Br}(B \to \rho / \omega \gamma)
= (1.8 ^{+0.6}_{-0.5} \pm 0.1) \times 10^{-6}~,
$$
has been reported at the Moriond conference \cite{Moriond}.
Implications of $B \to \rho \gamma$ and  $B \to \omega \gamma$
decays in the SM and the MSSM have been studied in the literature
\cite{alihandoko,alilunghi,kiers}.
The charged $B^\pm \to \rho^\pm \gamma$ decay mode provides clean signal
and has a branching ratio twice larger than that of
the neutral mode, by the isospin symmetry.
However, the long-distance (LD) effect on the charged mode
due to dominantly $W^\pm$-annihilation is very large ($\sim$30 \%),
which contaminates the CP violating effect \cite{alibraun,wyler}.
%Moreover, the charged mode is only related to the direct CP asymmetry
%and, therefore,  hardly applied to the extraction of $\beta$.
Consequently here we study the $B^0 \to \rho^0 \gamma$ decay
to explore the CP violation.
The time-dependent CP asymmetry in the neutral $B^0 \to \rho^0 \gamma$
decay probes the CP violation in the interference between decay and
mixing as well as the direct CP violation,
and it leads to the extraction of $\beta$ from both decay and mixing.
We note that the photon has two helicity states $\gamma_L$ and $\gamma_R$
which cannot be discriminated in the $B$ factory experiments.
Since CP violating asymmetry is defined when both $B$ and $\bar B$ mesons
decay into a same state, what we actually measure is the CP asymmetries
with the definite helicity in the final states.
In the SM, the operator which governs $b \to d \gamma$ decay is chiral
and the term with the conjugate operator is suppressed by $m_d/m_b$
and the CP asymmetry also suppressed accordingly.
Therefore, the new physics beyond the SM which involves a sizable
right-handed operator is required for a large time-dependent CP asymmetry
enough being observed in the experiment \cite{soni}.

In this letter, we consider the supersymmetric models which have
non-diagonal elements of the squark mass matrices
in the so-called super-CKM basis as a new physics model.
The non-diagonality of the down-type squark mass matrix
is parameterized by the mass insertions
$$(\delta_{ij})_{MN} \equiv (\tilde{m}_{ij}^2)_{MN} /{\tilde{m}}^2~,$$
where $\tilde{m}$ is the averaged squark mass.
Here $i$ and $j$ are flavor indices and $M$ and $N$ denote
chiralities, $L$ or $R$.
The $\delta$'s are complex in general and provide new CP phases.
To clarify and simplify our discussion,
we consider the $(\delta_{13})_{RL}$ and $(\delta_{13})_{RR}$
dominating cases
to produce the right-handed operators in the present work.
We note that the $(\delta_{13})_{LR}$ and $(\delta_{13})_{LL}$
mass insertions leads to the left-handed operators
and cannot generate the CP asymmetry only by themselves.
The $\Delta B = 1$ decay amplitude depends on
$(\delta_{13})_{RL,RR}$ linearly,
while the dependence of $M_{12}$ for the $\Delta B = 2$ process
is quadratic.
Then $b \rightarrow d\gamma$ decay would be more sensitive
to the SUSY contribution than the $B-\bar{B}$ mixing process
when the SUSY effects are smaller than the SM values as usual.
Thus we can observe the new physics effect through decay process
which is not observed through mixing induced processes.
In section II, we describe the $B^0 \to \rho^0 \gamma$ decay
and the time-dependent CP asymmetry.
The supersymmetric contributions are given in section III
and the numerical results given in the section IV.
We conclude in section V.

\section{CP asymmetry in $B^0 \to \rho^0 \gamma$ decay}

The relevant terms of the effective Hamiltonian
for the $b \to d \gamma$ decay is written as
\be
{\cal H}_{\rm eff} &=& \frac{4 G_F}{\sqrt{2}}
            \sum_{q=u,c} \left[ V_{qb} V_{qd}^*
                 \left( C_1 O_1^q + C_2 O_2^q
                 + C'_1 {O'_1}^q + C'_2 {O'_2}^q \right) \right.,
\nonumber \\
&&~~~~~~~~~~~
\left.
   - V_{tb} V_{td}^* \left( C^{\rm eff}_7 O_7
                 + C^{\prime \rm eff}_7 O'_7 \right)
            + \cdot \cdot \cdot \right],
\ee
where the operators are given by
\be
O_1^q &=& ({\bar{d}_L}^\alpha \gamma_\mu {q_L}^\beta)
        ({\bar{q}_L}^\beta \gamma^\mu {b_L}^\alpha),
\nonumber \\
O_2^q &=& (\bar{d}_L \gamma_\mu q_L)
             (\bar{q}_L \gamma^\mu  b_L),
\nonumber \\
O_7 &=& (e m_b/16 \pi^2) \bar{d}_L \sigma_{\mu \nu}
                                  F^{\mu \nu}  b_R,
\ee
and $O'_i$ are their complex conjugate operators.
The effective Wilson coefficient $ C^{(\prime) \rm eff}_7$ includes
the effects of operator mixing.

In the SM, $ C^{\prime \rm eff}_7$ is suppressed by the mass ratio
$m_d/m_b$ and so is the right polarized photon emission
$b_L \to q_R \gamma_R$.
The photon emission with the momentum $q$ and the polarization vector
$\epsilon_\mu (q)$ yields the electromagnetic field strength
$ F_{\mu \nu} = i (\epsilon_\nu q_\mu - \epsilon_\mu q_\nu)$.
The hadronic matrix element of $B \to \rho \gamma$ process
is parameterized in terms of two invariant form factors
$F_1^s$ and $F_2^s$
and the polarization vector of the $\rho$ meson, $\epsilon_\mu (k)$.
We can set $F_1^s = F_2^s \equiv F^s$ at $q^2 \sim 0$
for the real photon emission \cite{simma}.
We write the amplitudes for the final states of polarized photon as
\be
A_L &\equiv& \langle \rho \gamma_L | H_{\rm eff} | B^0 \rangle
= \frac{4 G_F}{\sqrt{2}} {C^{\prime \rm eff}_7}^* V^*_{tb} V_{td}
 \langle \rho \gamma_L | {O'_7}^\dagger | B^0 \rangle,
\nonumber \\
A_R &\equiv& \langle \rho \gamma_R | H_{\rm eff} | B^0 \rangle
= \frac{4 G_F}{\sqrt{2}} {C^{\rm eff}_7}^* V^*_{tb} V_{td}
 \langle \rho \gamma_R | {O_7}^\dagger | B^0 \rangle,
\nonumber \\
\bar{A}_L &\equiv& \langle \rho \gamma_L | H_{\rm eff} | \bar{B}^0 \rangle
= \frac{4 G_F}{\sqrt{2}} C^{\rm eff}_7 V_{tb} V^*_{td}
 \langle \rho \gamma_L | O_7 | \bar{B}^0 \rangle,
\nonumber \\
\bar{A}_R &\equiv& \langle \rho \gamma_R | H_{\rm eff} | \bar{B}^0 \rangle
= \frac{4 G_F}{\sqrt{2}} C^{\prime \rm eff}_7 V_{tb} V^*_{td}
 \langle \rho \gamma_R | O'_7 | \bar{B}^0 \rangle,
\ee
where
\be
\langle \rho(k) \gamma_L(q) | O_7 | \bar{B}^0(p) \rangle &=&
    \frac{e m_b}{16 \pi^2} \epsilon^\mu(q) \epsilon^\nu (k)
\left[  \epsilon_{\mu \nu \alpha \beta}  p^\alpha q^\beta
      - i (g_{\mu \nu} (q \cdot p) - p_\mu q_\nu )
\right] \cdot 2 F^s(q^2 \sim 0),
\nonumber \\
&=& \langle \rho \gamma_L | {O'_7}^\dagger | B^0 \rangle,
\nonumber \\
\langle \rho(k) \gamma_R(q) | O'_7 | \bar{B}^0(p) \rangle
&=& \frac{e m_b}{16 \pi^2} \epsilon^\mu(q) \epsilon^\nu (k)
\left[  \epsilon_{\mu \nu \alpha \beta}  p^\alpha q^\beta
      + i (g_{\mu \nu} (q \cdot p) - p_\mu q_\nu )
\right] \cdot 2 F^s(q^2 \sim 0),
\nonumber \\
&=& \langle \rho \gamma_R | {O_7}^\dagger | B^0 \rangle.
\ee
%Then the decay amplitude is given by \cite{alibraun}
%\be
%{\cal A}(B \to \rho \gamma) = -\frac{G_F}{\sqrt{2}} V_{tb} V_{td}^*
%        C^{\rm eff}_7
%    \frac{e m_b}{4 \pi^2} \epsilon^\mu(q) \epsilon_\mu (k)
%\left[  \epsilon_{\mu \nu \alpha \beta}  p^\alpha q^\beta
%      - i (g_{\mu \nu} (q \cdot p) - p_\mu q_\nu )
%\right] \cdot 2 F^s(q^2 \sim 0).
%\ee

For the neutral $B$ meson decay, the LD contribution
due to $W$-exchange is merely a few \% from the QCD sum rule
calculation \cite{alibraun,wyler}, so it will be ignored in our analysis.
We investigate the time-dependent CP asymmetry of neutral $B$ mesons,
which probes the direct CP violation and extracts the weak phase in mixing
and decay.
Since we cannot measure the photon helicity in practice,
the asymmetry which we will measure in the experiment is given by
\be
A_{\rm CP}(t) &=&
\frac{[\Gamma(\bar{B}_{\rm phys}^0 \to \bar{\rho}^0 \gamma_L)
      + \Gamma(\bar{B}_{\rm phys}^0 \to \bar{\rho}^0 \gamma_R) ]
      - [ \Gamma(B_{\rm phys}^0 \to \rho^0 \gamma_L)
      + \Gamma(B_{\rm phys}^0 \to \rho^0 \gamma_R) ] }
     {\Gamma(\bar{B}_{\rm phys}^0 \to \bar{\rho}^0 \gamma_L)
      + \Gamma(\bar{B}_{\rm phys}^0 \to \bar{\rho}^0 \gamma_R)
      + \Gamma(B_{\rm phys}^0 \to \rho^0 \gamma_L)
      + \Gamma(B_{\rm phys}^0 \to \rho^0 \gamma_R)  }
\nonumber \\
&\equiv& - {\cal C} \cos (\Delta m_B t) + {\cal S} \sin (\Delta m_B t),
\ee
where
\be
{\cal S} = \frac{ |A_L|^2 {\rm Im} \lambda_L + |A_R|^2 {\rm Im} \lambda_R}
           {|A_L|^2 + |A_R|^2},
\ee
with the parameter $\lambda_{L(R)}$ defined by
\be
\lambda_{L(R)} \equiv \sqrt{\frac{M_{12}^*}{M_{12}}}
                      \frac{\bar{A}_{L(R)}}{A_{L(R)}}.
\ee
Since $|A_L| = |\bar{A}_R|$ and $|A_R| = |\bar{A}_L|$,
the coefficient ${\cal C} = 0$ identically.
%Using the relation $\bar A_L/A_L = (|C_7|^2/|C'_7|^2)(\bar A_R/A_R)$,
%we obtain the familiar form
%$ {\cal S} =  2~{\rm Im} \lambda_R / (1 + |C'_7|^2/|C_7|^2)$.
The off-diagonal element $M_{12}$ describes the $B^0$--$\bar{B}^0$
mixing and $A_{L(R)}$ is the amplitude for
the $b \to d \gamma_{L(R)}$ decays.
We define
\be
\beta_{\rm mix} &=& \frac{1}{2} {\rm Arg} \left( M_{12} \right),
\nonumber \\
\beta_{\rm decay} &=& \frac{1}{2} {\rm Arg}
                          \left( \frac{{\bar A}_R}{A_R} \right)
 = \frac{1}{2} {\rm Arg} \left( \frac{{\bar A}_L}{A_L} \right),
\ee
and the coefficient ${\cal S}$ is expressed by
\be
{\cal S} = - \frac{2~|C_7||C'_7|}{|C_7|^2+|C'_7|^2}
          \sin (2 \beta_{\rm mix} - 2 \beta_{\rm decay}).
\ee
We can write
\be
2 \beta_{\rm decay} = 2 \beta_{\rm SM} + {\rm Arg}(C'_7) + {\rm Arg}(C_7).
\ee
Note that we have an additional factor
$2~|C_7||C'_7|/(|C_7|^2+|C'_7|^2)$,
which can enhance or suppress ${\cal S}$
by the new physics effect $|C'_7|$.
In the SM, $|C'_7|/|C_7| \sim {\cal O}(m_d/m_b)$
and the CP asymmetry is also suppressed by the corresponding factor.

\section{SUSY contributions}

We consider the gluino mediated penguin diagram
contribution to $b \to d \gamma$ decay in the MSSM.
By penguin diagrams with gluino-squark loop,
the Wilson coefficients $C'_i$ in the effective Hamiltonian of Eq. (2)
get contribution to produce $\gamma_R$ at the matching scale $\mu = m_W$.
After the RG evolution, we have
\be
C^{\rm eff}_7(m_b) &=& -0.31,
\nonumber \\
C^{\prime \rm eff}_7(m_b) &=&  \frac{\sqrt{2}}{G_F V_{tb} V_{td}^*}
       \left( 0.67~C^{\rm SUSY}_7(m_W) + 0.09~C^{\rm SUSY}_8(m_W) \right),
\ee
where the SUSY contributions are
\be
 C^{\rm SUSY}_7(m_W) &=& \frac{4 \alpha_s \pi Q_b}{3{\tilde m}^2}
         \left[ (\delta_{13})_{RR} M_4 (x)
              - (\delta_{13})_{RL} 4 B_1(x) \frac{m_{\tilde g}}{m_b} \right],
\nonumber \\
 C^{\rm SUSY}_8(m_W) &=& \frac{\alpha_s \pi}{6 {\tilde m}^2}
         \left[ (\delta_{13})_{RR} (9 M_3(x) - M_4(x) )
              + (\delta_{13})_{RL} \left(4 B_1(x)-9 \frac{B_2(x)}{x} \right)
                \frac{m_{\tilde g}}{m_b} \right],
\ee
with the squark mass insertions $(\delta_{13})_{RL}$ and $(\delta_{13})_{RR}$
and the squared mass ratio of the gluino mass to the average squark mass
$x=(m_{\tilde{g}}/\tilde{m})^2$
\cite{everett,baek,burascolangelo}.
Note that the SUSY contribution is more sensitive to $(\delta_{13})_{RL}$
than $(\delta_{13})_{RR}$ due to the enhancement factor $m_{\tilde g}/m_b$.
The loop functions $B_i(x)$ are found in the literature
\cite{everett,baek,burascolangelo}.
%The ratio of amplitudes is given by
%\be
%\frac{\bar{A}}{A} = \frac{{C^{\rm eff}_7}^*(m_b)}{C^{\rm eff}_7(m_b)}
% \frac{V^*_{tb} V_{td}}{V_{tb} V_{td}^*}.
%\ee
Since $\delta_{RL,RR}$ are complex in general,
the Wilson coefficients ${C'}_7^{\rm eff}(m_b)$ has nontrivial phase
which affects the phase of $\bar{A}/A$.

On the other hand, the $B$--$\bar{B}$ mixing is  affected
by the gluino-squark box diagrams in the MSSM.
Unlike the $\Delta B =1$ case,
the relevant  $\Delta B =2$ effective Hamiltonian
with the supersymmetric contribution contains new operators
which consist of scalar-scalar interactions,
\be
O'_{S2} &=& (\bar{d}_\alpha (1+\gamma_5) b_\alpha)
      (\bar{d}_\beta  (1+\gamma_5) b_\beta),
\nonumber \\
O'_{S3} &=& (\bar{d}_\alpha (1+\gamma_5) b_\beta)
      (\bar{d}_\alpha  (1+\gamma_5) b_\beta),
\ee
when we introduce only the RL and RR mass insertions.
The Wilson coefficients $C_{1,S2,S3}$
corresponding to the SM operators $O_1 = (\bar{d} \gamma_\mu (1-\gamma_5) b)
(\bar{d} \gamma_\mu (1-\gamma_5) b)$ and
$O_{S2,S3} = O'_{S2,S3} (L \leftrightarrow R)$ consist of
the SM part and the supersymmetric contributions,
while $C'_{S2}$ and $C'_{S3}$
corresponding to the above operators are entirely supersymmetric.
Their explicit expression at the scale $\mu = M_{\rm SUSY}$
can be found in Refs. \cite{gabrielli,ko}.
Ignoring the RG running effects between $M_{\rm SUSY}$ and $m_W$,
we perform the RG evolution from  $m_W$ to $m_b$ scale
to obtain the evolved Wilson coefficients.
The SUSY contributions are given by
\be
{C^{(\prime)}}_i^{\rm SUSY} (m_b) = \sum_k \sum_j
       \left( b_k^{(i,j)} + \eta c_k^{(i,j)} \right)
             \eta^{a_k} {C^{(\prime)}}_j^{\rm SUSY}(M_{\rm SUSY}),
\ee
where the magic numbers $a_k$, $b_k^{(i,j)}$ and $ c_k^{(i,j)}$
are found in  Ref. \cite{BB}
and $\eta = \alpha_s(M_{SUSY})/\alpha_s(m_W)$.
The off-diagonal element $M_{12}$  obtained by
$ M_{12} =
\langle B^0|{\cal H}^{\Delta B=2}_{eff}|\bar{B}^0 \rangle /2 m_B$
consists of the bag parameters $B_i$ and the decay constant $f_{B_d}$
in vacuum insertion approximation.

\section{Numerical results}

We write
$(\delta_{13})_{RL(RR)} \equiv |(\delta_{13})_{RL(RR)}| e^{i \varphi}$.
Figure 1 shows the allowed values of the quantity ${\cal S}$
as a function of the phase $\varphi$,
assuming $(\delta_{13})_{RL}$ dominating case with $|(\delta_{13})_{RL}|=0.001$.
We vary the weak phase $\gamma$ from 0 to $ 2\pi$.
Hereafter we use the input parameters as follows:
$ m_B= 5.3~{\rm GeV}$, $ m_t= 174.3~{\rm GeV}$, $ m_b= 4.6~{\rm GeV}$,
and $ \alpha_s(m_Z) = 0.118$.
The decay constant $f_{B_d} = 200 \pm 30$ MeV is
the main source of the theoretical uncertainty
and the bag parameters are those of  Ref. \cite{bag}
$B_1 = 0.87,~~ B_2=0.82,~~ B_3=1.02$.
The supersymmetric scale is taken to be
$m_{\tilde{g}} \approx \tilde{m} \approx M_{\rm SUSY} \approx 500$ GeV.
We require that the mass difference $\Delta m_B$
and $\beta_{\rm mix}$ in $B \to J/\psi K$ decay
should be within the experimental limit:
$\Delta m_B = 0.489 \pm 0.008$ ps$^{-1}$ \cite{pdb}
and $\sin 2 \beta_{\rm mix} = 0.734 \pm 0.055$ \cite{sin2beta}.
On the other hand, it is reported that
the branching ratio of the inclusive $B \to X_d \gamma$ decay
%puts a stringent constraint on the RL mass insertion
puts a stringent constraint on the flavor mixing parameters
of the gluino-squark contributions \cite{ko}.
In spite of the recent measurement of the decay rate
of the exclusive $B \to \rho / \omega \gamma$ channel,
we do not use  ${\rm Br}(B \to \rho / \omega \gamma)$
as a constraint here since it involves a large theoretical uncertainty
in the form factor.
Instead, we assume a moderate upper bound
${\rm Br}(B \to X_d \gamma) < 1.0 \times 10^{-5}$ in this analysis,
following Ref. \cite{ko} and referring to the measurement
of the exclusive channel \cite{Moriond},
although no experimental limit on the branching ratio
for the inclusive decay is given yet.
In the figure, the black region corresponds to the allowed values for
the phase of $(\delta_{13})_{RL}$,
%while the grey (green) region denotes the parameter set which satisfies
while the grey (green) region to the parameter sets which satisfy
the $\Delta m_B$ and $\sin 2 \beta_{\rm mix}$ constraints
but exceeds the bound on the branching ratio of
the inclusive $B \to X_d \gamma$ decay.
We find that large CP violating asymmetry reaching $\sim \pm 60 \%$,
is possible
while all other experimental constraints are satisfied.
As we can notice from Eq. (10), the maximal value of the CP violating asymmetry
is set by the coefficient $2~|C_7||C'_7|/(|C_7|^2+|C'_7|^2)$.
In the present case, this coefficient is roughly 0.6 because
the allowed value of $|C'_7|$ is about 0.1 for $|(\delta_{13})_{RL}| = 0.001$
(note $C_7=C_7^{SM}=-0.31$ for $(\delta_{13})_{RL(RR)}$ dominating cases).
%If we assume larger values for $|(\delta_{13})_{RL}|$,
%allowed region for values of $\sin (2\beta_{\rm decay}+2\beta_{\rm mix})$
If we increase the assumed value of $|(\delta_{13})_{RL}|$,
the allowed range for the CP asymmetry, ${\cal S}$
becomes expanded and may reach even $\pm 100\%$ if $|C_7|=|C'_7|$,
which is realized when $|(\delta_{13})_{RL}| \sim 0.003$.
However, such large $|C'_7|$ values would give large
$B \to X_d \gamma$ branching ratio
which exceeds the assumed upper limit of the branching ratio.
Still the CP asymmetry ${\cal S} \sim \pm 0.8$ is allowed
while satisfying experimental constraints including
the inclusive branching ratio bound.

The plot for the allowed range of ${\cal S}$ with respect to
$|(\delta_{13})_{RL}|$
is depicted in  Fig. 2
when the phase $\varphi$ is fixed to be zero.
The black region and the grey (green) region
are defined as  in Fig. 1.
We see that the value of the CP asymmetry is zero
for $|(\delta_{13})_{RL}| = 0$,
which corresponds to the SM case with vanishing $d$- quark mass limit.
As already explained, when $|(\delta_{13})_{RL}|$ value increase,
the CP asymmetry  ${\cal S}$ increase and reach its maximal value at
$|(\delta_{13})_{RL}| \sim 0.003 $ and then decrease again.
Here we notice two points.
First, $|(\delta_{13})_{RL}|$ is strongly constrained by the
inclusive branching ratio.
In fact, the branching ratio bound is much stronger than
the bounds on $\Delta m_B$ and $\sin 2 \beta_{\rm mix}$.
This is because $b \to d \gamma$ decay rate is very sensitive to
$(\delta_{13})_{RL}$ mass insertion due to the enhancement factor $m_{\tilde{g}}/m_b$
as one can check it from Eq. (13). On the other hand, there is no such enhancement factor
for the $B$--$\bar{B}$ mixing process.
Second, a large CP violation (up to $\sim 35\%$) is still possible
even when the only source of CP violating phase is the CKM mixing matrix,
$i.e,$ when ${C'}_7^{\rm eff}(m_b)$ is real.
The branching ratio ${\rm Br}(B \to X_d \gamma)$
and CP asymmetry ${\cal S}$ provide the complimentary information
on  $(\delta_{13})_{RL}$.

%Figure 3 and 4 are corresponding to Fig. 1 and 2
Figure 3 and 4 are the counterparts of Fig. 1 and 2 respectively,
when $(\delta_{13})_{RR}$ dominates.
The black region and the grey (green) region
are defined as in Fig. 1 and 2.
In Fig. 3,  $| (\delta_{13})_{RR} | = 0.03$ is assumed
while the phase $\varphi$ is varied from 0 to $2\pi$.
We find that CP asymmetry ${\cal S}$ reach $\sim \pm 5\%$ at best.
This small CP asymmetry comes from the fact that $|C'_7|$ is small,
compared to $|C_7|$. The CP asymmetry increases a little
as we increase $|(\delta_{13})_{RR}|$.
But we cannot increase $|(\delta_{13})_{RR}|$ enough to produce
large CP asymmetries (say, larger than $10\%$).
This is because the $\Delta m_B$ and $\sin 2 \beta_{\rm mix}$ constraints
are stronger than the branching ratio bounds
in the $(\delta_{13})_{RR}$ dominating case.
(Note that the opposite is true for the $(\delta_{13})_{RR}$ dominating case.)
The phase $\varphi$ is set to be zero in Fig. 4,
while varying $|(\delta_{13})_{RR}|$.
We can see that the CP asymmetry is so small that
it might be difficult to measure it in the near future.

In short, large CP asymmetry is possible
for the $(\delta_{13})_{RL}$ dominating case.
On the other hand, only small CP symmetry is allowed
for the $(\delta_{13})_{RR}$ dominating case.
The difference of these two cases is due to the enhancement factor
$m_{\tilde{g}}/m_b$
in the case of $(\delta_{13})_{RL}$ mass insertion.
%We can see that effect of $(\delta_{13})_{RR}$
%is milder than that of $(\delta_{13})_{RL}$
%due to the enhancement factor $m_{\tilde{g}}/m_b$.

\section{Concluding remarks}

If we observe a sizable CP asymmetry in $B^0 \to \rho^0 \gamma$ decay,
it will be a clear evidence of the new physics
beyond the SM.
Although it is hardly expected that the time dependent CP asymmetry
of $B^0 \to \rho^0 \gamma$ will be measured in the present $B$-factory,
it will be achieved in the next generation of $B$-factory
with about 100 times more $B$ mesons produced.
Due to the agreement of the SM prediction with
the present $\Delta m_B$ data and the CP asymmetry
in $B \to J/\psi K$ decay,
we favor the new physics which contributes less
to the $B$--$\bar{B}$ mixing
but has a strong  effect on the $b \to d \gamma$
penguin diagram.
In this paper, we showed that the RL mass insertion of squark mixing
of the MSSM can produce a large CP asymmetry
of $B^0 \to \rho^0 \gamma$ decay process.
\\

%In conclusion, we suggested to explore $\beta_{\rm decay}$
%in $b \to d \gamma$ decay as a good test of the validity of the SM
%and found that $\beta_{\rm decay}$ with the $(\delta_{13})_{RL}$
%squark mass insertion in the MSSM can
%have a value very different  from the SM prediction.
%We anticipate to measure $B^0 \to \rho^0 \gamma$ process
%and extract $\sin (2\beta_{\rm decay}+2\beta_{\rm mix})$
%in the next generation of super $B$-factories.

\acknowledgments

The work of C.S.K. was supported
by Grant No. R02-2003-000-10050-0 from BRP of the KOSEF.
The work of YGK was supported by the Korean Federation
of Science and Technology Societies through
the Brain Pool program.
This work was supported by Korea Research Foundation Grant
(KRF-2003-050-C00003, KYL).
We would like to thank G. Cvetic and P. Ko for their helpful comments.

%%%%%%%%%%%%%%%%%% References
%%%%%%%%%%%%%%%%%%%%%%%%%%%%%%%%%%%%%%%%%%%%%%%%%%%%%%%
\def\PRD #1 #2 #3 {Phys. Rev. D {\bf#1},\ #2 (#3)}
\def\PRL #1 #2 #3 {Phys. Rev. Lett. {\bf#1},\ #2 (#3)}
\def\PLB #1 #2 #3 {Phys. Lett. B {\bf#1},\ #2 (#3)}
\def\NPB #1 #2 #3 {Nucl. Phys. {\bf B#1},\ #2 (#3)}
\def\ZPC #1 #2 #3 {Z. Phys. C {\bf#1},\ #2 (#3)}
\def\EPJ #1 #2 #3 {Euro. Phys. J. C {\bf#1},\ #2 (#3)}
\def\JHEP #1 #2 #3 {JHEP {\bf#1},\ #2 (#3)}
\def\NC #1 #2 #3 {Nuovo Cimento {\bf#1A},\ #2 (#3)}
\def\IJMP #1 #2 #3 {Int. J. Mod. Phys. A {\bf#1},\ #2 (#3)}
\def\MPL #1 #2 #3 {Mod. Phys. Lett. A {\bf#1},\ #2 (#3)}
\def\PTP #1 #2 #3 {Prog. Theor. Phys. {\bf#1},\ #2 (#3)}
\def\PR #1 #2 #3 {Phys. Rep. {\bf#1},\ #2 (#3)}
\def\RMP #1 #2 #3 {Rev. Mod. Phys. {\bf#1},\ #2 (#3)}
\def\PRold #1 #2 #3 {Phys. Rev. {\bf#1},\ #2 (#3)}
\def\IBID #1 #2 #3 {{\it ibid.} {\bf#1},\ #2 (#3)}

\begin{center}
\begin{figure}[htb]
%\vskip 1cm
\hbox to\textwidth{\hss\epsfig{file=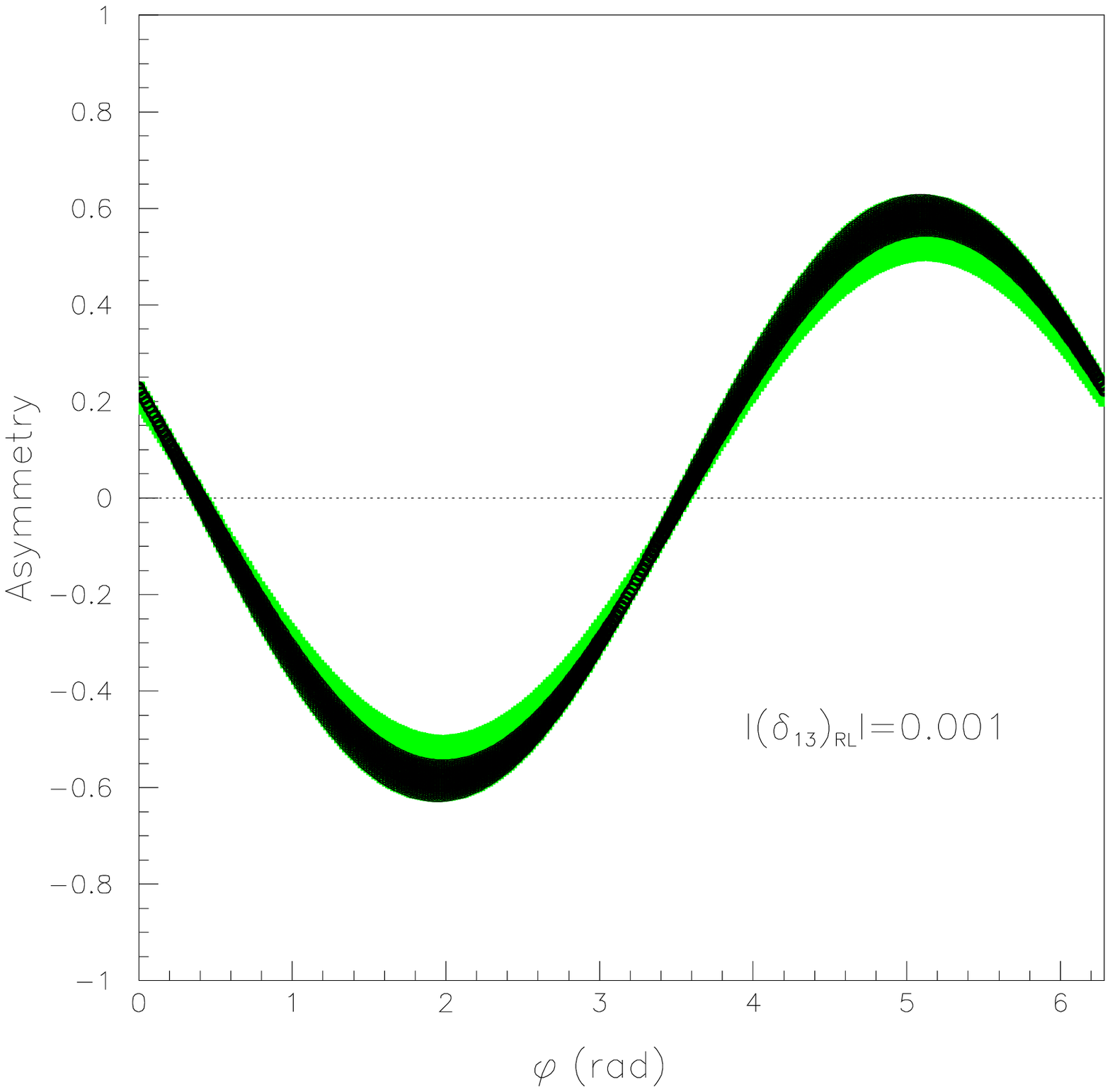,width=16cm}\hss}
 \vskip -1.5cm
\vspace{1cm}
\caption{
The time-dependent CP asymmetry ${\cal S}$
as a function of the phase of $(\delta_{13})_{RL}$.
$|(\delta_{13})_{RL}| =0.001$ is assumed.
The black region denotes allowed points while grey (green) region
excluded points by the inclusive $b \to d \gamma$
branching ratio bound.
}
\end{figure}
\end{center}

\newpage

\begin{center}
\begin{figure}[htb]
%\vskip 1cm
\hbox to\textwidth{\hss\epsfig{file=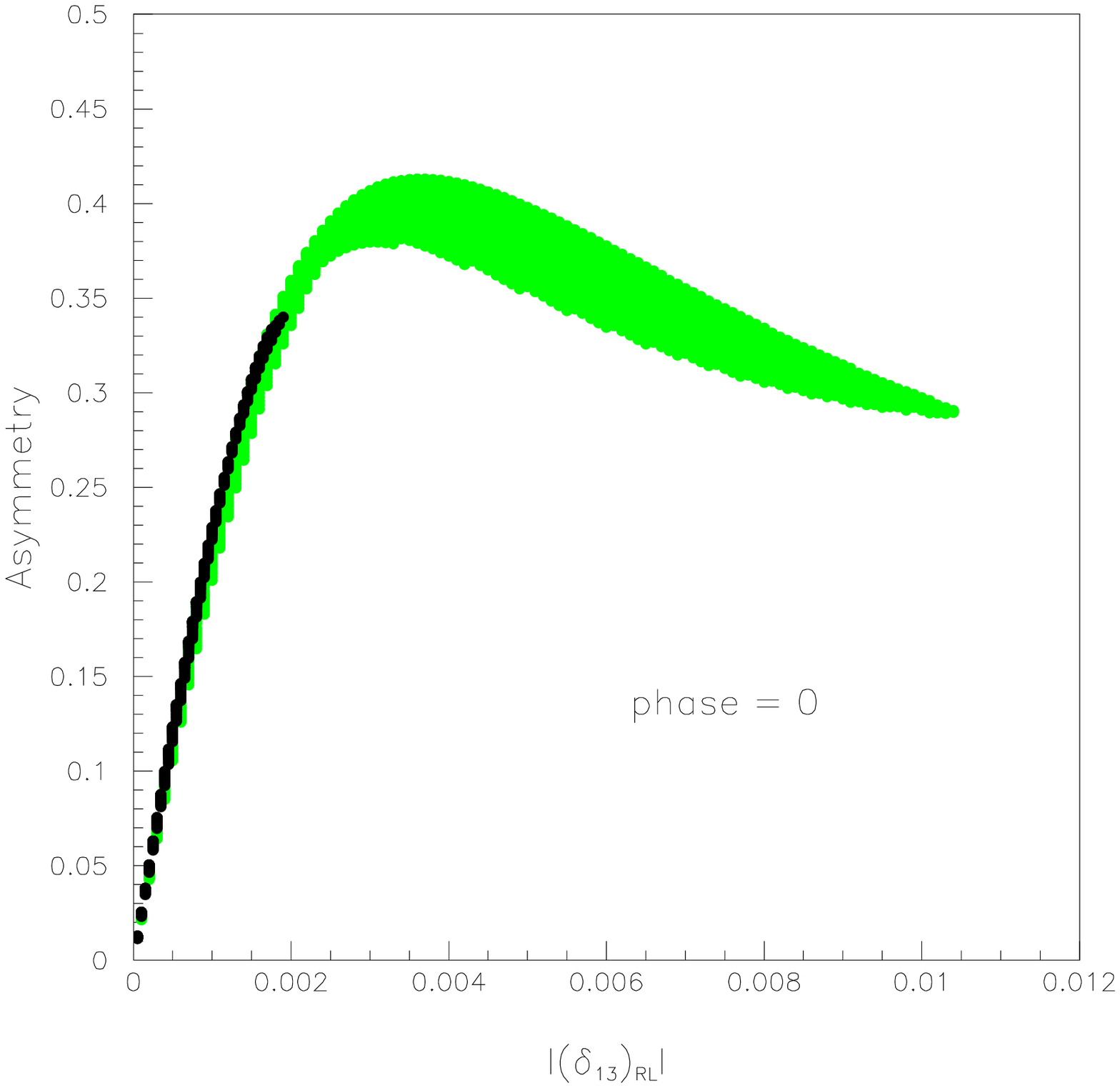,width=16cm}\hss}
 \vskip -1.5cm
\vspace{1cm}
\caption{
The time-dependent CP asymmetry ${\cal S}$
as a function of $|(\delta_{13})_{RL}|$.
The phase of $(\delta_{13})_{RL}$ is assumed to be 0.
The black region denotes allowed points while grey (green) region
excluded points by the inclusive $b \to d \gamma$
branching ratio bound.
}
\end{figure}
\end{center}

\newpage

\begin{center}
\begin{figure}[htb]
%\vskip 1cm
\hbox to\textwidth{\hss\epsfig{file=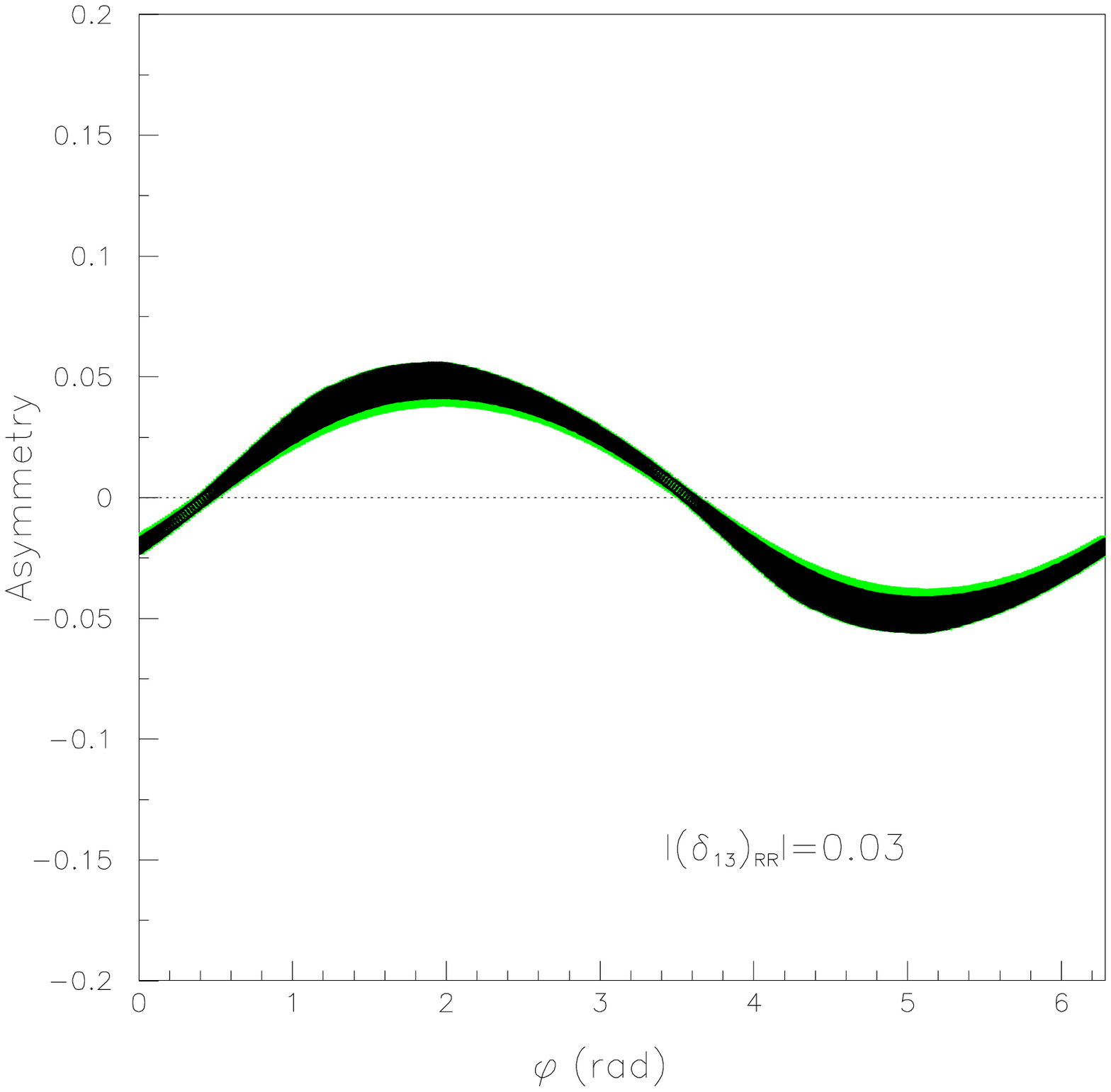,width=16cm}\hss}
 \vskip -1.5cm
\vspace{1cm}
\caption{
The time-dependent CP asymmetry ${\cal S}$
as a function of the phase of $(\delta_{13})_{RR}$.
$|(\delta_{13})_{RR}| =0.03$ is assumed.
The black region denotes allowed points while grey (green) region
excluded points by the inclusive $b \to d \gamma$
branching ratio bound.
}
\end{figure}
\end{center}

\newpage

\begin{center}
\begin{figure}[htb]
%\vskip 1cm
\hbox to\textwidth{\hss\epsfig{file=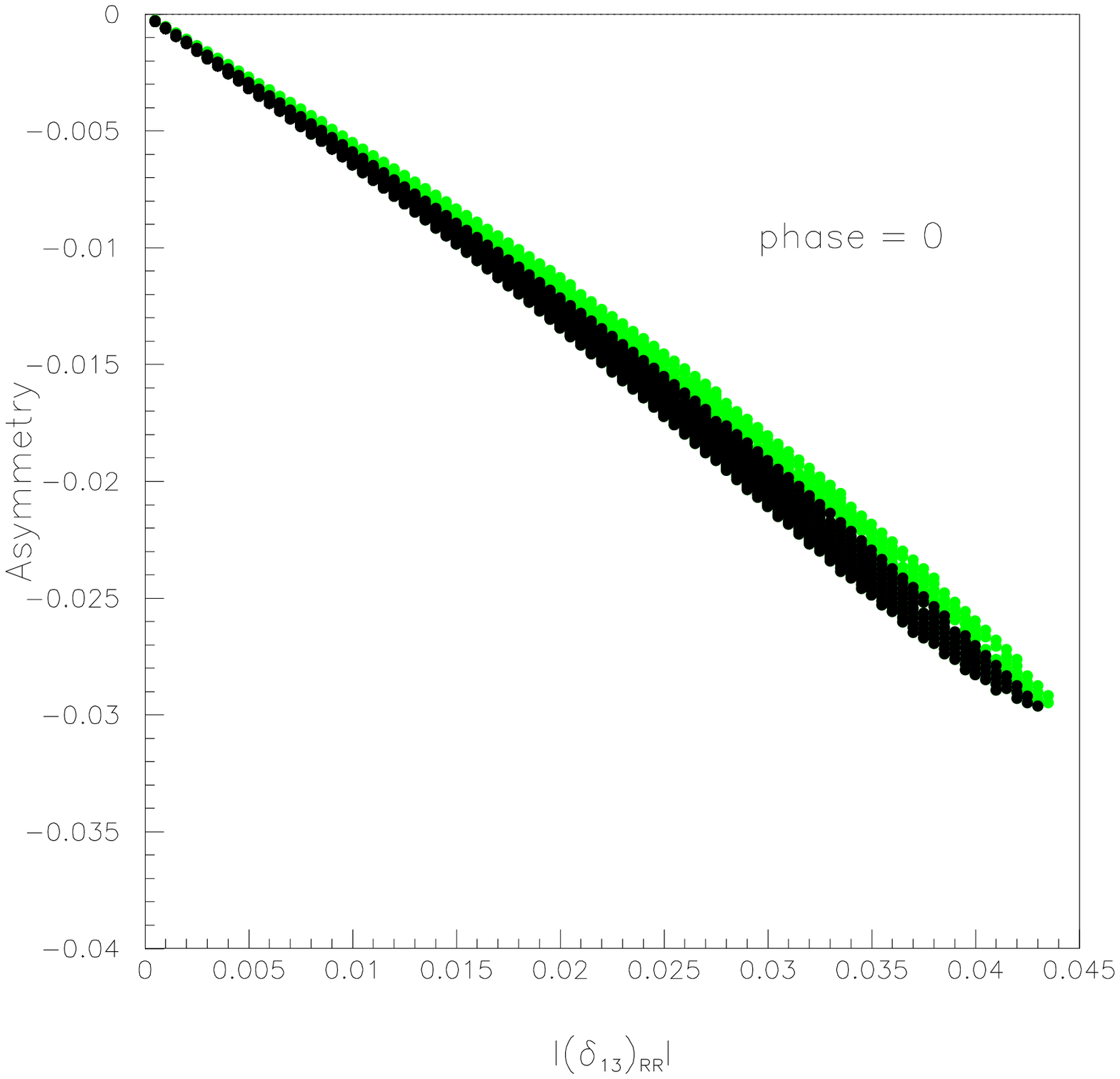,width=16cm}\hss}
 \vskip -1.5cm
\vspace{1cm}
\caption{
The time-dependent CP asymmetry ${\cal S}$
as a function of $|(\delta_{13})_{RR}|$.
The phase of $(\delta_{13})_{RR}$ is assumed to be 0.
The black region denotes allowed points while grey (green) region
excluded points by the inclusive $b \to d \gamma$
branching ratio bound.
}
\end{figure}
\end{center}

%\end{multicols}
%\vfil\eject

\end{document}